\documentclass[prd,amsmath,amssymb,preprintnumbers,superscriptaddress,twocolumn,10pt,nofootinbib]{revtex4-1} 

\usepackage{graphicx}
\usepackage{subfigure}
\usepackage{float}

\usepackage{dcolumn}
\usepackage{bm}
\usepackage{amssymb}
\usepackage{latexsym}
\usepackage{booktabs}
\usepackage{amsmath}
\usepackage{multirow}
\usepackage{url}
\usepackage{changes}
\usepackage{soul} 
\usepackage{color, xcolor}
\usepackage[colorlinks=true, linkcolor=red, citecolor=blue]{hyperref}

\hypersetup{
    colorlinks=true,
    linkcolor=red,
    citecolor=blue
}
\IfFileExists{orcidlink.sty}{\usepackage{orcidlink}}{}

\providecommand{\orcidlink}[1]{%
  \href{https://orcid.org/#1}{%
    \includegraphics[height=1.6ex]{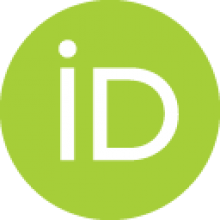}%
  }%
}

\def\red#1{{\textcolor{red}{#1}}}

\usepackage[normalem]{ulem}
\usepackage{array}
\usepackage{enumerate}

\def\be{\begin{equation}}
\def\ee{\end{equation}}

\def\bea{\begin{eqnarray}}
\def\eea{\end{eqnarray}}
\newcommand{\bes}{\begin{equation*}}
\newcommand{\ees}{\end{equation*}}
\newcommand{\beqa}{\begin{eqnarray}}
\newcommand{\eeqa}{\end{eqnarray}}

\begin{document}

\title{Multi-band cross-correlation dark sirens: enhancing cosmological parameter and gravitational-wave bias constraints}

\author{Ji-Yu Song\orcidlink{0009-0003-8111-0470}}
\affiliation{Liaoning Key Laboratory of Cosmology and Astrophysics, College of Sciences, Northeastern University, Shenyang 110819, China}

\author{Ya-Nan Du\orcidlink{0009-0003-0453-9046}}
\affiliation{Liaoning Key Laboratory of Cosmology and Astrophysics, College of Sciences, Northeastern University, Shenyang 110819, China}

\author{Yue-Yan Dong\orcidlink{0009-0003-9108-4974}}
\affiliation{Liaoning Key Laboratory of Cosmology and Astrophysics, College of Sciences, Northeastern University, Shenyang 110819, China}

\author{Jing-Fei Zhang\orcidlink{0000-0002-3512-2804}}
\affiliation{Liaoning Key Laboratory of Cosmology and Astrophysics, College of Sciences, Northeastern University, Shenyang 110819, China}

\author{Xin Zhang\orcidlink{0000-0002-6029-1933}}\thanks{Corresponding author: \href{mailto:zhangxin@neu.edu.cn}{zhangxin@neu.edu.cn}}
\affiliation{Liaoning Key Laboratory of Cosmology and Astrophysics, College of Sciences, Northeastern University, Shenyang 110819, China}
\affiliation{MOE Key Laboratory of Data Analytics and Optimization for Smart Industry, Northeastern University, Shenyang 110819, China}
\affiliation{National Frontiers Science Center for Industrial Intelligence and Systems Optimization, Northeastern University, Shenyang 110819, China}

\begin{abstract}

Multi-band gravitational-wave (GW) observation, combining space-borne and ground-based detectors across different frequency bands, can improve the sky localization of compact binary sources by two to three orders of magnitude compared to single-band detection. This enhancement is crucial for cross-correlation dark siren analyses, since the sky localization uncertainty directly determines the noise level of the GW angular power spectrum. In this work, we present the first Fisher forecast for cross-correlation dark siren cosmology with multi-band GW observations, cross-correlating GW events from the Einstein Telescope (ET), Cosmic Explorer (CE), and B-DECIGO with the Chinese Space-station Survey Telescope photometric galaxy survey. We compare three network configurations: the multi-band B-DECIGO+ET+2CE (BDET2CE), the ground-only ET+2CE (ET2CE), and B-DECIGO alone. In the $\Lambda$CDM model, BDET2CE achieves $\sigma(h)/h = 0.35\%$, improving by $37\%$ over the ground-only ET2CE ($0.55\%$) and by $86\%$ over B-DECIGO alone ($2.45\%$). Extending to the $w_0w_a$CDM framework, the multi-band advantage on cosmological parameters becomes more moderate, with BDET2CE improving $\sigma(h)/h$ by $\sim4\%$ over ET2CE and $\sim22\%$ over B-DECIGO. The most striking advantage of multi-band observation lies in the per-bin measurement of the GW clustering bias $b_{\rm GW}(z)$: at $z \sim 1$--$2$, BDET2CE constrains the bias to $\sim3\%$ precision, compared to $\sim$8--$60\%$ for ET2CE and $\sim$20--$33\%$ for B-DECIGO. These precise, redshift-resolved bias measurements open a new avenue for probing the astrophysics of compact binary mergers, enabling constraints on formation channels such as isolated binary evolution and dynamical assembly that predict distinct clustering signatures.

\end{abstract}



\maketitle

\section{Introduction}\label{sec:Introduction}

The Hubble tension, now reaching a ${\sim}6\sigma$ discrepancy between the locally measured Hubble constant from the JWST and HST Cepheid-calibrated distance ladder and the value inferred from the Planck CMB observations assuming the $\Lambda$CDM model \cite{Riess:2025chq, Planck:2018vyg}, has become one of the most pressing open problems in modern cosmology. The tension persists across independent late-universe probes, suggesting that it may point to physics beyond the standard cosmological model rather than systematic errors in any single measurement \cite{Vagnozzi:2019ezj,Knox:2019rjx,Verde:2019ivm,DiValentino:2021izs,Perivolaropoulos:2021jda, Schoneberg:2021qvd,Abdalla:2022yfr, Kamionkowski:2022pkx,CosmoVerseNetwork:2025alb}.

Adding to this puzzle, the Dark Energy Spectroscopic Instrument (DESI) has recently released its Data Release 2 (DR2) results based on baryon acoustic oscillation measurements from over 14 million galaxies and quasars spanning redshifts up to $z \sim 4$ \cite{DESI:2024mwx, DESI:2025zgx}. The combination of DESI DR2 with CMB data yields an approximately $3\sigma$ preference for dynamical dark energy over the cosmological constant in the $w_0w_a$CDM parameterization. This preference is maintained or strengthened when different Type Ia supernova datasets are included, with the statistical significance ranging from 2.8 to $4.2\sigma$ depending on the supernova sample \cite{DESI:2025zgx}. Importantly, the favored dark energy evolution with $w_0 > -1$ and $w_a < 0$, in which $w$ transitions from below $-1$ in the past to above $-1$ at present, does not alleviate the Hubble tension but rather exacerbates it \cite{DESI:2024mwx, DESI:2025zgx}. These developments underscore the need for independent late-universe probes of the cosmic expansion history with entirely different systematics, which can provide complementary constraints on both the Hubble constant and the dark energy equation of state.

Gravitational-wave (GW) standard sirens offer an independent late-universe probe of the cosmic expansion \cite{Schutz:1986gp, Holz:2005df}. The waveform amplitude of GW signals from compact binary coalescences is inversely proportional to the luminosity distance $d_{\mathrm{L}}$, yielding an absolute distance measurement that requires no external calibration \cite{DelPozzo:2011vcw, Chen:2017rfc}. Combined with redshift information of the GW source, this enables direct constraints on the expansion history. The landmark event GW170817, with its electromagnetic counterpart, provided the first standard-siren measurement of $H_0$ \cite{LIGOScientific:2017adf, LIGOScientific:2017vwq, LIGOScientific:2017apx, LIGOScientific:2017pwl}. However, such bright sirens with identified host galaxies are exceedingly rare; even in the era of the third-generation (3G) ground-based detectors such as the Einstein Telescope (ET) \cite{Punturo:2010zz} and Cosmic Explorer (CE) \cite{LIGOScientific:2016wof}, they are expected to constitute less than 1\% of all detections \cite{Han:2023exn, Han:2024sxm, Han:2025fii}. The vast majority of GW events are dark sirens lacking electromagnetic counterparts, for which redshift inference requires alternative methods \cite{Taylor:2011fs,Gray:2019ksv,Mastrogiovanni:2021wsd,Ezquiaga:2022zkx,Mastrogiovanni:2023emh,Gray:2023wgj}. See refs.~\cite{Zhao:2010sz, Cai:2016sby, Wang:2018lun, Zhang:2018byx, Yang:2019vni, Zhang:2019ylr, Zhang:2019loq, Wang:2019tto, Zhao:2019gyk, Jin:2020hmc, Zhao:2020ole, Jin:2021pcv, Qi:2021iic, Wang:2021srv, Jin:2022tdf, Jin:2022qnj, Wu:2022dgy, Song:2022siz, Wang:2022oou, Jin:2023tou, Jin:2023sfc, Jin:2023zhi, Yu:2023ico, Li:2023gtu, Feng:2024lzh, Feng:2024mfx, Xiao:2024nmi, Feng:2025wbz, Xiao:2025mcg, Song:2025bio, Song:2025ddm, Dong:2025ikq, Zhang:2025yhi, Jin:2025dvf, Li:2025vuh, Song:2026kii, Zhu:2026gsu, Zhao:2026uvn, Das:2026glo,Sun:2026dga, Liu:2026dug, Dong:2026uxr, Agapito:2026kak, McMahon:2026nhi} for applications of standard sirens in cosmology.

Beyond measuring distances, GW sources detected by future observatories will be sufficiently numerous and well-localized to serve as tracers of the large-scale structure of the universe \cite{Raccanelli:2016cud, Scelfo:2018sny, Vijaykumar:2020pzn, Libanore:2020fim}. Angular cross-correlation between GW source catalogs and galaxy surveys provides a powerful framework for extracting cosmological information from dark sirens \cite{Oguri:2016dgk, Mukherjee:2019wcg, Mukherjee:2020hyn, Bera:2020jhx, Mukherjee:2022afz, Bosi:2023amu, Diaz:2021pem, Ghosh:2023ksl, Pedrotti:2025tfg}. In this approach, GW events binned by luminosity distance are cross-correlated with galaxies binned by redshift; only the correct distance-redshift relation yields the maximum cross-correlation signal, thereby constraining cosmological parameters. Unlike event-by-event host-galaxy searches, this statistical method is largely immune to galaxy catalog incompleteness and population-model assumptions. In our companion paper \cite{Du:2025odq}, we performed such an analysis for the $\Lambda$CDM model, cross-correlating the Chinese Space-station Survey Telescope (CSST) \cite{CSST:2025ssq} photometric galaxy catalog with GW events from 3G ground-based detectors, and demonstrated the significant potential of this synergy for constraining $H_0$ and $\Omega_m$.

In addition to cosmological parameters, the cross-correlation simultaneously measures the GW clustering bias $b_{\rm GW}$, which quantifies the relation between the GW source distribution and the underlying dark matter density field. The GW bias carries rich astrophysical information: different formation channels for compact binary mergers, including isolated binary stellar evolution, dynamical assembly in dense stellar environments, active galactic nucleus disk-assisted mergers, and primordial black hole scenarios, predict distinct clustering signatures \cite{Scelfo:2020jyw,Libanore:2023ovr, Chakravarti:2026bcv}. At the same time, the GW bias enters the angular power spectrum amplitude and must be simultaneously constrained with cosmological parameters. Measuring the GW bias across redshift is therefore important both for understanding compact binary formation and for ensuring the robustness of cross-correlation cosmological analyses. An important factor affecting both the cosmological and bias constraints is the sky localization uncertainty of GW sources, which enters the noise term of the angular power spectrum and directly impacts the constraining power.

Multi-band GW observation, which combines space-borne deci-hertz detectors with ground-based hertz-band detectors, has the potential to dramatically improve the sky localization of GW sources. The proposed baseline Decihertz Interferometer Gravitational-wave Observatory (B-DECIGO) mission \cite{Kawamura:2020pcg}, operating in the 0.1--10~Hz band, would observe compact binary inspirals months to years before their mergers enter the ET/CE sensitivity band. The joint multi-band analysis can improve sky localization by two to three orders of magnitude compared to single-band observations \cite{Cutler:2019krq, Zhao:2023ilw,Zhu:2021bpp,Dong:2024bvw}. Given this improvement, we are motivated to ask: can multi-band GW observations provide a significant advantage for GW$\times$galaxy cross-correlation analyses, particularly in constraining dark energy, the Hubble constant, and the GW clustering bias?

In this work, we address this question by performing a comprehensive Fisher forecast for the angular power spectrum cross-correlation between multi-band GW observations and the CSST photometric galaxy survey within both the $\Lambda$CDM and $w_0w_a$CDM frameworks. We compare three detector network configurations: the multi-band B-DECIGO + ET + two CEs (BDET2CE), the ground-only ET + two CEs (ET2CE), and B-DECIGO alone. To ensure conservative and robust results, we adopt a per-bin bias parameterization in which the galaxy and GW clustering biases are treated as independent free parameters in each redshift bin.

This paper is organized as follows. In sect.~\ref{sec:method}, we present the angular power spectrum formalism for cross-correlating GW sources and galaxies, including the Fisher matrix methodology. Sect.~\ref{sec:simulation} describes the modeling of the CSST photometric survey and the simulation of GW observations for the three detector configurations. In sect.~\ref{sec:results}, we present the parameter constraints and discuss the multi-band advantage. Finally, sect.~\ref{sec:conclusions} summarizes the main findings.

\section{Methods}\label{sec:method}

We quantify the clustering of galaxies and GW sources through tomographic angular power spectra, following the formalism of ref.~\cite{Pedrotti:2025tfg}. All calculations are performed with the publicly available code \texttt{pyccl}\footnote{\url{https://ccl.readthedocs.io/en/latest/}} \cite{LSSTDarkEnergyScience:2018yem}, using number-count tracers under the Limber approximation. We consider two cosmological frameworks: the $\Lambda$CDM model, in which the dark energy equation of state (EoS) is a constant $w = -1$, and its extension $w_0w_a$CDM, which parameterizes the dark energy EoS as $w(a) = w_0 + w_a(1-a)$ with fiducial values $w_0 = -1$ and $w_a = 0$. For the $w_0w_a$CDM analysis, the parameterized post-Friedmann prescription \cite{Fang:2008sn} is employed to ensure numerical stability when the equation of state crosses the phantom divide $w = -1$ during finite-difference calculations. Since GW observations measure luminosity distances rather than redshifts, we bin GW events by their measured $d_{\mathrm{L}}$ and construct the GW tracer kernel in redshift space using the cosmology-dependent $d_{\mathrm{L}}(z)$ relation. Specifically, for each GW luminosity-distance bin $[d_{{\rm L},j}^{\min}, d_{{\rm L},j}^{\max}]$, we compute the effective redshift distribution as
\begin{equation}\label{eq:nz_eff}
\frac{\mathrm{d}N_{\rm GW}^{\rm eff}}{\mathrm{d}z}\bigg|_j = \frac{\mathrm{d}N_{\rm GW}}{\mathrm{d}z}(z) \times W^{\rm GW}[d_{\mathrm{L}}(z,\lambda),\, d_{{\mathrm{L}},j}],
\end{equation}
where $\mathrm{d}N_{\rm GW}/\mathrm{d}z$ is the intrinsic merger rate distribution (cosmology-independent, estimated from the mock catalog via kernel density estimation), and $W^{\rm GW}$ is the luminosity-distance window function defined in eq.~\eqref{eq:9}. The key feature of this construction is that $d_{\mathrm{L}}(z,\lambda)$ depends on the cosmological parameters $\lambda$: when computing Fisher derivatives, each perturbed cosmology produces a different $d_{\mathrm{L}}(z)$ mapping, which shifts the effective $z$-range of the GW kernel and changes the overlap with the galaxy redshift bins. This sensitivity to the distance-redshift relation is the primary mechanism through which the cross-correlation constrains cosmological parameters \cite{Pedrotti:2025tfg}. The \texttt{pyccl} number-count tracer is then constructed with this effective $\mathrm{d}N_{\rm GW}^{\rm eff}/\mathrm{d}z$ for each $d_{\mathrm{L}}$-bin at each cosmology evaluation.

\subsection{Angular power spectra of galaxies and GW sources}\label{ssec:multi-tracer}

Both galaxies and GW sources trace the underlying dark matter distribution through their spatial clustering. For a tracer $X$ (either galaxies or GW sources), the number density contrast at radial coordinate $x$ in direction $\boldsymbol{n}$ is defined as
\begin{equation}\label{eq:1}
\Delta^X(\boldsymbol{n},x)=\frac{N^X(\boldsymbol{n},x)-\langle N^X\rangle(x)}{\langle N^X\rangle(x)},   
\end{equation}
where $\langle\cdots\rangle$ denotes the ensemble average. Decomposing $\Delta^X$ into spherical harmonics yields
\begin{equation}\label{eq:2}
\Delta^X(\boldsymbol{n},x)=\sum_{\ell=0}^\infty\sum_{m=-\ell}^\ell s_{\ell m}^X(x)Y_{\ell m}(\boldsymbol{n}),
\end{equation}
with harmonic coefficients $s_{\ell m}^X(x)$. To extract clustering information, we divide the galaxy sample into $N_{\rm g}$ redshift bins and the GW sample into $N_{\rm GW}$ luminosity-distance bins, computing the tomographic angular power spectrum $C_\ell^{XY}(x_i, x_j)$ between bin pairs via
\begin{equation}\label{eq:3}
\langle s_{\ell m}^{X}(x_{i})s_{\ell^{\prime}m^{\prime}}^{Y^{*}}(x_{j})\rangle=\delta_{\ell\ell^{\prime}}\delta_{mm^{\prime}}C_{\ell}^{XY}(x_{i},x_{j}),
\end{equation}
where $^*$ is the complex conjugate, $X = Y$ gives auto-correlations, and $X \neq Y$ gives cross-correlations. Within perturbation theory and adopting weighted radial bins, the angular power spectrum takes the form
\begin{equation}\label{eq:4}
C_\ell^{XY}(x_i,x_j)=\frac{2}{\pi}\int\mathrm{d}kk^2P(k)\Delta_\ell^{X,x_i}(k)\Delta_\ell^{Y,x_j}(k),
\end{equation}
{where $P(k)$ is the linear matter power spectrum and $\Delta_\ell^{X,x_i}(k)$ is the effective transfer function of tracer $X$ in bin $i$:}
\begin{equation}\label{eq:5}
\Delta_\ell^{X,x_i}(k)=\int_0^\infty\mathrm{d}xw^X(x,x_i)\Delta_\ell^X(k,x),
\end{equation}
Here $\Delta_\ell^X(k,x)$ encodes the Fourier-space number density fluctuations, receiving contributions from density, velocity (including redshift-space distortions, RSDs, and Doppler effects), lensing, and gravitational potential terms \cite{Bonvin:2011bg, Challinor:2011bk, Fonseca:2023uay}. For GW tracers in luminosity-distance space, RSDs are replaced by luminosity-distance space distortions (LSDs) \cite{Zhang:2018nea, Vijaykumar:2020pzn, Libanore:2020fim}. The full transfer function decomposes as
\begin{equation}\label{eq:6}
\begin{aligned}
    \Delta_{\ell}^{X}(k,x)=&\Delta_\ell^{X,\mathrm{den}}(k,x)+\Delta_\ell^{X,\mathrm{vel}}(k,x)\\&+\Delta_\ell^{X,\mathrm{len}}(k,x)+\Delta_\ell^{X,\mathrm{gr}}(k,x),
\end{aligned}
\end{equation}
with the velocity component further splitting into
\begin{equation}\label{eq:7}
\Delta_\ell^{X,\mathrm{vel}}(k,x)=\Delta_\ell^{X,\mathrm{RSD/LSD}}(k,x)+\Delta_\ell^{X,\mathrm{dop}}(k,x),
\end{equation}
Following ref.~\cite{Pedrotti:2025tfg}, we retain only the dominant contributions: density, lensing, and RSD/LSD. The Doppler and gravitational potential terms are subdominant on the sub-horizon scales probed by our multipole range, suppressed by factors of $\mathcal{H}/(kc)$ relative to the RSD contribution \cite{Bonvin:2011bg, Challinor:2011bk}. Neglecting them is standard practice in angular power spectrum forecasts for both galaxy \cite{Euclid:2024yrr} and GW surveys \cite{Scelfo:2020jyw, Pedrotti:2025tfg, Libanore:2020fim}.

The weighted window function $w^X(x, x_i)$ in eq.~\eqref{eq:5} is constructed from an unweighted window $W^X(x, x_i)$ and the observed number density distribution:
\begin{equation}\label{eq:8}
w^{X}(x,x_{i})=W^{X}(x,x_{i})\frac{\mathrm{d}N_{\mathrm{obs}}^{X}}{\mathrm{d}x}\frac{1}{\int\mathrm{d}x^{\prime}W^{X}(x^{\prime},x_{i})\frac{\mathrm{d}N_{\mathrm{obs}}^{X}}{\mathrm{d}x^{\prime}}}.
\end{equation}

The unweighted window function convolves a top-hat bin selection with a log-normal measurement error model:
\begin{equation}\label{eq:9}
\begin{aligned}
W^X(x,x_i) & =\int_0^\infty\operatorname{d}yS^X(y,x_i)\mathcal{L}(y|x) \\
           & =\frac{1}{2}\left\{\operatorname{erf}[u(x^{i+1},x)]-\operatorname{erf}[u(x^i,x)]\right\},
\end{aligned}
\end{equation}
with
\begin{equation}\label{eq:10}
u(y,x)=\frac{\ln y-\ln x}{\sqrt{2}\sigma_{\ln x}},
\end{equation}
where $\sigma_{\ln x}$ is the relative measurement error (in redshift or luminosity distance), and $x^i$, $x^{i+1}$ are the bin boundaries. The observed number density $\mathrm{d}N_{\rm obs}^X / \mathrm{d}x$ for galaxies and GW sources is modeled in sect.~\ref{sec:simulation}. We divide the CSST galaxy sample into 15 equal-number photometric redshift bins spanning $z \in [0, 4]$. The GW luminosity-distance bins are obtained by mapping these redshift boundaries to $d_{\mathrm{L}}$ at the fiducial cosmology; once defined, these $d_{\mathrm{L}}$-bin boundaries remain fixed (cosmology-independent), while the corresponding $z$-range of each GW tracer kernel varies with cosmology through the $d_{\mathrm{L}}(z,\lambda)$ relation as described above.

Under the Limber approximation \cite{Bellomo:2020pnw, Bernal:2020pwq}, the line-of-sight integration simplifies, and we retain the density, lensing, and RSD/LSD contributions (see Appendix A of ref.~\cite{Pedrotti:2025tfg} for full expressions). The dominant density-density terms read
\begin{equation}\label{eq:11}
{
\begin{aligned}
C_{{\ell},\mathrm{den}}^{\mathrm{gGW}}(z_{i},d_{\mathrm{L},j})
=& \int_{0}^{\infty}\mathrm{d}z\,w^{\mathrm{g}}(z,z_{i})\,w^{\mathrm{GW}}[d_{\rm{L}}(z,\lambda),d_{\mathrm{L},j}]\\
&\times\frac{\mathrm{d}d_{\mathrm{L}}}{\mathrm{d}z}(z,\lambda) \frac{H(z)}{cr(z)^{2}}b_{\mathrm{GW}}(z)b_{\mathrm{g}}(z)\\
&\times P\left[\frac{\ell+1/2}{r(z,\lambda)},z\right],
\end{aligned}
}
\end{equation}
\begin{equation}
{
\begin{aligned}
C_{{\ell},\mathrm{den}}^{\mathrm{gg}} (z_{i},z_{j}) 
=& \int_{0}^{\infty}\mathrm{d}z\,w^{\mathrm{g}}(z,z_{i})\,w^{\mathrm{g}}(z,z_{j})\frac{H(z)}{cr(z)^{2}}\\
&\times\left[b_{\mathrm{g}}(z)\right]^2\,P\left[\frac{\ell+1/2}{r(z,\lambda)},z\right],
\end{aligned}
}
\end{equation}
\begin{equation}
{
\begin{aligned}
C_{{\ell},\mathrm{den}}^{\mathrm{GWGW}}(d_{\mathrm{L},i},d_{\mathrm{L},j})
=& \int_{0}^{\infty}\mathrm{d}z\,w^{\mathrm{GW}}[d_{\rm{L}}(z,\lambda),d_{\mathrm{L},i}]\\
&\times{w^{\mathrm{GW}}[d_{\rm{L}}(z,\lambda),d_{\mathrm{L},j}]}\\
&\times\left[\frac{\mathrm{d}d_{\mathrm{L}}}{\mathrm{d}z}(z,\lambda)\right]^2\frac{H(z)}{cr(z)^{2}}\\
&\times\left[b_{\mathrm{GW}}(z)\right]^2\,P\left[\frac{\ell+1/2}{r(z,\lambda)},z\right],
\end{aligned}
}
\end{equation}
where $r(z,\lambda)$ is the comoving distance and $\lambda$ denotes the cosmological parameters governing the distance-redshift relation. The galaxy window $w^{\rm g}$ operates in redshift space while the GW window $w^{\rm GW}$ operates in luminosity-distance space; their radial overlap is maximized only when the assumed cosmology matches the true one \cite{Pedrotti:2025tfg}. This sensitivity to the distance-redshift relation is the physical basis for constraining cosmological parameters through the cross-correlation amplitude.

The clustering bias $b_X$ of tracer $X$ relates its number density contrast to the underlying dark matter overdensity through $b_X = \delta_X / \delta_{\rm m}$. On the large scales relevant to this analysis, $b_X$ depends only on redshift and is independent of scale. In contrast to approaches that assume a global parametric form for the bias redshift evolution, we adopt a per-bin parameterization in which each tomographic bin has an independent bias parameter. Specifically, we introduce scaling factors $\alpha_{\rm g}^{(i)}$ and $\alpha_{\rm GW}^{(i)}$ for each bin $i$, such that the effective bias in bin $i$ is
\begin{equation}
b_{\rm g}^{(i)}(z) = \alpha_{\rm g}^{(i)} \, b_{\rm g}^{\rm fid}(z), \qquad b_{\rm GW}^{(i)}(z) = \alpha_{\rm GW}^{(i)} \, b_{\rm GW}^{\rm fid}(z),
\end{equation}
where the fiducial bias models are $b_{\rm g}^{\rm fid}(z) = b_0(1+z)^{b_1}$ with $b_0 = 1$, $b_1 = 1$ \cite{Gong:2019yxt}, and $b_{\rm GW}^{\rm fid}(z) = A_{\rm GW}(1+z)^{\gamma}$ with $A_{\rm GW} = 1.20$, $\gamma = 0.59$ \cite{Peron:2023zae}. All scaling factors have fiducial values $\alpha^{(i)} = 1$ and are treated as free parameters in the Fisher analysis. This yields 15 independent galaxy bias parameters and 15 independent GW bias parameters, which are marginalized over when deriving cosmological constraints. This per-bin approach avoids the assumption of a specific functional form for the bias evolution and produces more conservative, robust constraints.

The lensing contribution introduces a magnification bias $s_X(z)$, which quantifies the change in the observed number of sources due to gravitational lensing effects. Following ref.~\cite{Pedrotti:2025tfg}, we parameterize it as a cubic polynomial $s(z) = s_0 + s_1 z + s_2 z^2 + s_3 z^3$, with $(s_0, s_1, s_2, s_3) = (0.0842, 0.0532, 0.298, -0.0113)$ for the galaxy survey and $(-0.00559, 0.0292, 0.00344, 0.00258)$ for the GW survey.

\subsection{Noise and the range of angular scales}\label{ssec:noise}

\begin{table*}
\centering
\caption{Multipole ranges for each redshift/$d_{\mathrm{L}}$ bin. $\ell_{\min}$ and $\ell_{\max}^{\rm NL}$ are set by the linear regime and are common to all configurations. $\ell_{\max}^{\rm LOC}$ and $\ell_{\rm damp}$ depend on GW sky localization and differ dramatically between detectors. For BDET2CE, $\ell_{\max}^{\rm LOC} \gg \ell_{\max}^{\rm NL}$ in all bins, so the effective multipole range for GW spectra is entirely set by the nonlinear scale. For ET2CE, $\ell_{\rm damp}$ falls below $\ell_{\max}^{\rm NL}$ at $z \gtrsim 0.5$, severely damping the GW signal and limiting the cross-correlation information (see sect.~\ref{ssec:gwbias}).\label{tab:ell}}
\renewcommand{\arraystretch}{1.2}
\begin{tabular*}{\textwidth}{@{\extracolsep{\fill}}ccccccccc}
\hline\hline
 & & & \multicolumn{2}{c}{BDET2CE} & \multicolumn{2}{c}{ET2CE} & \multicolumn{2}{c}{B-DECIGO} \\
\cmidrule(lr){4-5} \cmidrule(lr){6-7} \cmidrule(lr){8-9}
Bin ($z$) & $\ell_{\min}$ & $\ell_{\max}^{\rm NL}$ & $\ell_{\max}^{\rm LOC}$ & $\ell_{\rm damp}$ & $\ell_{\max}^{\rm LOC}$ & $\ell_{\rm damp}$ & $\ell_{\max}^{\rm LOC}$ & $\ell_{\rm damp}$ \\
\hline
$0.00$--$0.25$ & 5 & 55 & 161300 & 30941 & 4204 & 907 & 156773 & 21952 \\
$0.25$--$0.35$ & 5 & 104 & 41628 & 14385 & 1331 & 489 & 40892 & 11231 \\
$0.35$--$0.43$ & 5 & 140 & 42954 & 11698 & 1173 & 372 & 41856 & 8720 \\
$0.43$--$0.51$ & 5 & 174 & 29554 & 9774 & 872 & 307 & 28576 & 7397 \\
$0.51$--$0.58$ & 10 & 207 & 20497 & 8096 & 757 & 250 & 19492 & 5936 \\
$0.58$--$0.65$ & 10 & 240 & 18487 & 6481 & 563 & 205 & 18338 & 4939 \\
$0.65$--$0.73$ & 10 & 277 & 18349 & 5750 & 512 & 172 & 18142 & 4671 \\
$0.73$--$0.81$ & 10 & 319 & 17174 & 5225 & 374 & 161 & 17352 & 4275 \\
$0.81$--$0.90$ & 15 & 366 & 12463 & 4479 & 364 & 130 & 12382 & 3796 \\
$0.90$--$1.00$ & 15 & 421 & 12258 & 3907 & 278 & 109 & 12673 & 3668 \\
$1.00$--$1.11$ & 15 & 485 & 10567 & 3611 & 256 & 99 & 11092 & 3413 \\
$1.11$--$1.25$ & 15 & 567 & 8714 & 3089 & 244 & 89 & 9604 & 3029 \\
$1.25$--$1.44$ & 20 & 683 & 7629 & 2685 & 189 & 76 & 8538 & 2785 \\
$1.44$--$1.74$ & 20 & 871 & 6564 & 2277 & 152 & 61 & 7478 & 2428 \\
$1.74$--$4.00$ & 20 & 1410 & 4578 & 1628 & 99 & 40 & 6023 & 1896 \\
\hline\hline
\end{tabular*}
\end{table*}


The observed angular power spectrum includes both the theoretical signal and a shot-noise component. Expanding the noise into spherical harmonics $n_{\ell m}(x_i)$, the total observed coefficients are
\begin{equation}\label{eq:12}
a_{\ell m}^X(x_i)=s_{\ell m}(x_i)+n_{\ell m}(x_i).
\end{equation}

The noise power spectrum satisfies
\begin{equation}\label{eq:13}
\langle n_{\ell m}^X(x_i)n_{\ell^{\prime}m^{\prime}}^{Y^*}(x_j)\rangle=\delta_{\ell\ell^{\prime}}\delta_{mm^{\prime}}\delta_{XY}\delta_{ij}\mathcal{N}^X(x_i),
\end{equation}
where the shot noise $\mathcal{N}^X(x_i)$ is the inverse of the angular number density in bin $i$:
\begin{equation}\label{eq:14}
\mathcal{N}^X(x_i)=\left[\int_0^\infty\mathrm{d}xW^X(x,x_i)\frac{\mathrm{d}^2N_{\mathrm{obs}}^X}{\mathrm{d}x\mathrm{d}\Omega}\right]^{-1}.
\end{equation}

Assuming uncorrelated signal and noise, the observed power spectrum becomes
\begin{equation}\label{eq:15}
\langle a_{\ell m}^X(x_i),a_{\ell^{\prime}m^{\prime}}^{Y^*}(x_j)\rangle=\delta_{\ell\ell^{\prime}}\delta_{mm^{\prime}}\tilde{C}_\ell^{XY}(x_i,x_j),
\end{equation}
\begin{equation}\label{eq:16}
\tilde{C}_\ell^{XY}(x_i,x_j)\equiv C_\ell^{XY}(x_i,x_j)+\delta_{XY}\delta_{ij}\mathcal{N}^X(x_i).
\end{equation}

Sky localization uncertainties of GW sources further suppress the angular power spectra at high multipoles through a Gaussian damping:
\begin{equation}\label{eq:17}
C_\ell^{\mathrm{gGW}}(d_{\mathrm{L},i},z_j)\mapsto C_\ell^{\mathrm{gGW}}(d_{\mathrm{L},i},z_j)\times e^{-\ell(\ell+1)/\ell_{\mathrm{damp}}^2},
\end{equation}
\begin{equation}
\begin{aligned}\label{eq:18}
    C_\ell^{\mathrm{GWGW}}(d_{\mathrm{L},i},d_{\mathrm{L},j})\mapsto & C_\ell^{\mathrm{GWGW}}(d_{\mathrm{L},i},d_{\mathrm{L},j})\\&\times e^{-2\ell(\ell+1)/\ell_{\mathrm{damp}}^2},
\end{aligned}
\end{equation}
\begin{equation}\label{eq:19}
    \ell_{\mathrm{damp}}^2=\frac{(2\pi)^{3/2}}{\Delta\Omega_{1\sigma}},
\end{equation}
where $\Delta\Omega_{1\sigma}$ is the median ($Q_{50}$) of the $1\sigma$ (68\% CL) sky localization areas within each bin, following the standard 2D Gaussian beam convention \cite{Libanore:2020fim, Pedrotti:2025tfg}.

The accessible multipole range in each bin is bounded from above by two considerations. The nonlinear cutoff is set by
\begin{equation}\label{eq:20}
    \ell_{\max}^{\mathrm{NL}} = r(z_i) k_{i,\mathrm{max}},
\end{equation}
where $r(z_i)$ is the comoving distance, $k_{i,\mathrm{max}} = \pi/(2R_i)$ with $\sigma^2(R_i) = 0.25$ defining the linear-nonlinear transition \cite{Pedrotti:2025tfg}, and both quantities are evaluated at the $n(z)$-weighted effective redshift $z_{\rm eff}^{(i)} = \int z\,n(z)\,\mathrm{d}z / \int n(z)\,\mathrm{d}z$ within each bin. This choice is particularly important for the widest bin ($z = 1.74$--$4.0$), where the arithmetic midpoint $z_{\rm mid} = 2.87$ would overestimate the nonlinear cutoff; the $n(z)$-weighted $z_{\rm eff} = 2.13$ yields a more conservative and physically representative value. The angular resolution limit imposed by GW localization is
\begin{equation}\label{eq:21}
    \ell_{\max}^{\mathrm{LOC}} = \pi / \sqrt{\Delta\Omega_{\min}},
\end{equation}
where $\Delta\Omega_{\min}$ is the 2nd percentile of the 90\% CL sky localization areas. For galaxy auto-correlations, $\ell_{\max} = \ell_{\max}^{\rm NL}$; for spectra involving GW tracers, $\ell_{\max} = \min(\ell_{\max}^{\rm NL}, \ell_{\max}^{\rm LOC})$. The minimum multipole follows ref.~\cite{Tanidis:2019teo}: $\ell_{\min} = 5$ for $z < 0.5$, $10$ for $0.5 < z < 0.75$, $15$ for $0.75 < z < 1.25$, and $20$ for $z > 1.25$. All adopted values are listed in Table~\ref{tab:ell}. Notably, for the BDET2CE multi-band configuration, $\ell_{\max}^{\rm LOC}$ exceeds $\ell_{\max}^{\rm NL}$ by one to two orders of magnitude in all bins (see Table~\ref{tab:ell}), so the effective multipole range is entirely set by the nonlinear scale rather than by localization. This is a direct consequence of the dramatically improved sky localization achieved through multi-band observation.

\begin{figure*}
    \centering
    \includegraphics[width=0.85\linewidth]{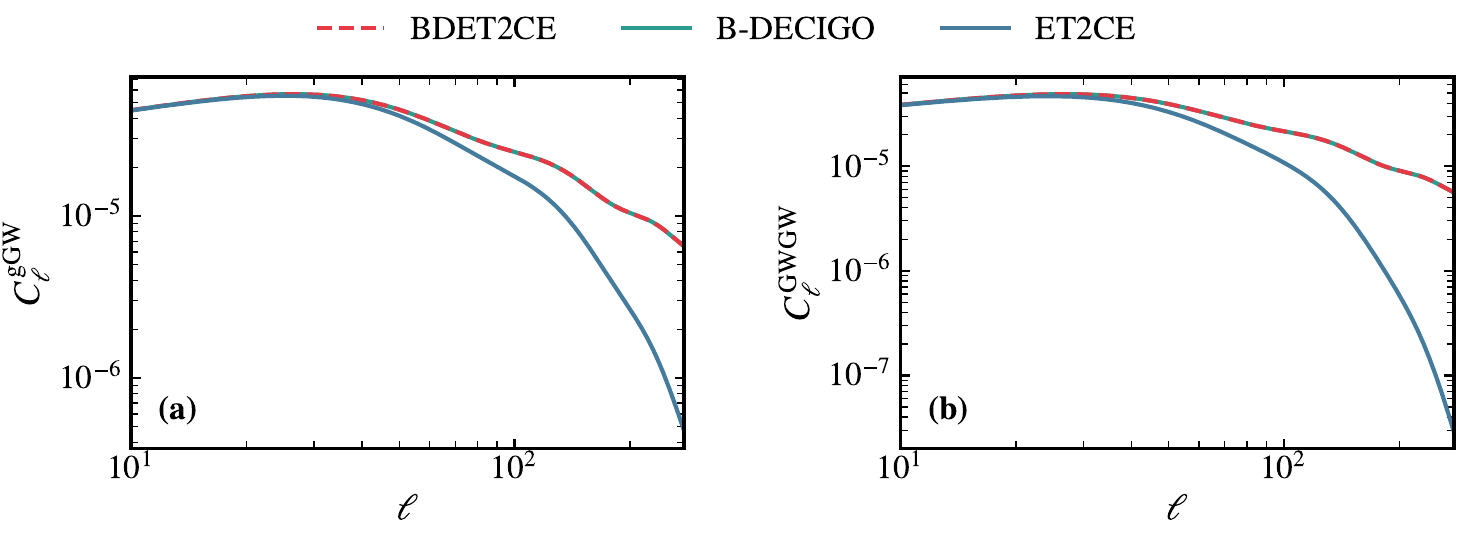}
    \caption{Effect of sky localization damping on the galaxy-GW cross-correlation spectrum $C_\ell^{\rm gGW}$ (left) and the GW auto-correlation spectrum $C_\ell^{\rm GWGW}$ (right) in the redshift bin $z \in [0.65, 0.73]$ with $T_{\rm obs} = 10$~yr. The multi-band BDET2CE configuration (red) preserves the signal across the full multipole range, while the ground-only ET2CE (blue) suffers severe damping at $\ell \gtrsim 100$ due to its larger sky localization uncertainties. The space-only B-DECIGO (green) experiences relatively mild damping due to its high-precision sky localization, despite its smaller event catalog.}
    \label{fig:damping}
\end{figure*}

Figure~\ref{fig:damping} illustrates the impact of sky localization damping on the angular power spectra for the three detector configurations in the redshift bin $z \in [0.65, 0.73]$. For BDET2CE, the damping is negligible across the entire accessible multipole range, whereas for ET2CE the signal is suppressed by orders of magnitude at $\ell \gtrsim 100$. B-DECIGO falls between the two, benefiting from its superior single-source localization but limited by a smaller event catalog. This comparison highlights the advantage of multi-band observations in preserving the cross-correlation signal at high multipoles.

\begin{figure}
    \centering
    \includegraphics[width=\columnwidth]{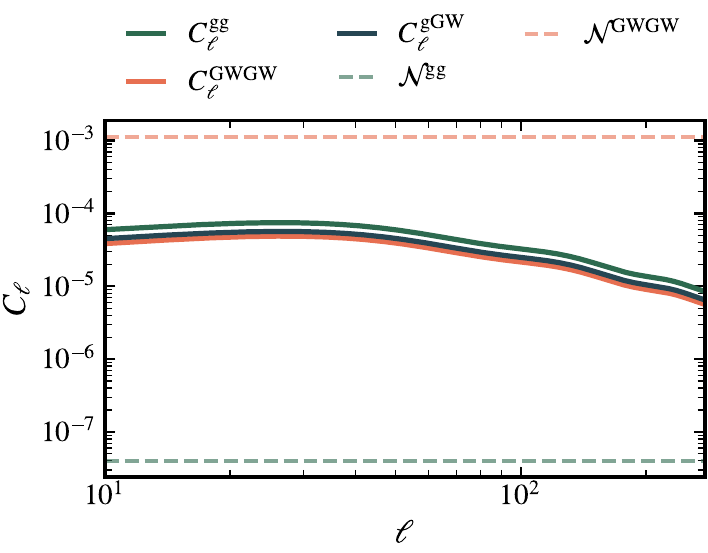}
    \caption{Angular power spectra and noise levels in the redshift bin $z \in [0.65, 0.73]$ for BDET2CE with $T_{\rm obs} = 10$~yr. The galaxy auto-correlation $C_\ell^{\rm gg}$ (green) exceeds its shot noise $\mathcal{N}^{\rm gg}$ (green dashed) by three to four orders of magnitude, while the GW auto-correlation $C_\ell^{\rm GWGW}$ (orange) lies well below its shot noise $\mathcal{N}^{\rm GWGW}$ (orange dashed). The galaxy-GW cross-correlation $C_\ell^{\rm gGW}$ (dark blue) has no shot noise contribution.}
    \label{fig:Cl}
\end{figure}

Figure~\ref{fig:Cl} shows the angular power spectra and noise levels for the same redshift bin, illustrating the signal hierarchy: the galaxy auto-correlation $C_\ell^{\rm gg}$ is signal-dominated across the entire multipole range, while the GW auto-correlation $C_\ell^{\rm GWGW}$ lies well below its shot noise due to the much smaller number of GW events compared to galaxies. The galaxy-GW cross-correlation $C_\ell^{\rm gGW}$ is free of shot noise, since shot noise arises from the self-Poisson fluctuation of discrete tracers and vanishes identically for cross-correlations between two distinct populations ($N_\ell^{XY} \propto \delta_{XY}$). This is the key advantage of the cross-correlation approach: although the GW auto-correlation is noise-dominated by itself, the cross-correlation extracts the clustering signal without shot-noise contamination. Note, however, that the measurement uncertainty of the cross-correlation still depends on the total power spectra of both tracers, including their respective shot-noise contributions, through the covariance matrix. This is why the sky localization of GW sources, which determines the effective number density and thus the GW shot noise, remains a critical factor for the constraining power.

\subsection{Fisher information matrix}\label{ssec:fisher matrix}

We forecast parameter constraints using the Fisher information matrix. For a parameter set $\{\theta_\alpha\}$, the matrix elements are
\begin{equation}
\begin{aligned}
    F_{ij}&=-\left\langle\frac{\partial^2\ln\mathcal{L}}{\partial\theta_\alpha\partial\theta_\beta}\right\rangle\\&=f_\mathrm{sky}\sum_\ell\frac{2\ell+1}{2}\mathrm{Tr}[\mathcal{C}_\ell^{-1}(\partial_\alpha\mathcal{C}_\ell)\mathcal{C}_\ell^{-1}(\partial_\beta\mathcal{C}_\ell)],
\end{aligned}
\end{equation}
where $f_{\mathrm{sky}} = \Omega_{\mathrm{sky}} / (4\pi)$ is the sky coverage fraction. $\mathcal{C}_\ell$ is the covariance matrix. Assuming the galaxy survey is divided into $N$ tomographic redshift bins $z_1$, ..., $z_N$, and the GW survey is divided into $M$ luminosity-distance bins $d_{\mathrm{L},1}$, ..., $d_{\mathrm{L},M}$, the symmetric covariance matrix can be expressed as:
\begin{equation}
\begin{aligned}
    &\mathcal{C}_\ell =\\
&\resizebox{\columnwidth}{!}{$
\begin{pmatrix}
\tilde{C}_\ell^{\rm gg}(z_1,z_1) & \cdots & \tilde{C}_\ell^{\rm gg}(z_1,z_N)
 & \tilde{C}_\ell^{\rm gGW}(z_1,d_{\mathrm{L},1}) & \cdots 
 & \tilde{C}_\ell^{\rm gGW}(z_1,d_{\mathrm{L},M}) \\[4pt]
 & \ddots & \vdots & \vdots & & \vdots \\[4pt]
 & & \tilde{C}_\ell^{\rm gg}(z_N,z_N)
 & \tilde{C}_\ell^{\rm gGW}(z_N,d_{\mathrm{L},1}) & \cdots 
 & \tilde{C}_\ell^{\rm gGW}(z_N,d_{\mathrm{L},M}) \\[4pt]
 & & & \tilde{C}_\ell^{\rm GWGW}(d_{\mathrm{L},1},d_{\mathrm{L},1}) 
 & \cdots & \tilde{C}_\ell^{\rm GWGW}(d_{\mathrm{L},1},d_{\mathrm{L},M}) \\[4pt]
 & & & & \ddots & \vdots \\[4pt]
 & & & & & \tilde{C}_\ell^{\rm GWGW}(d_{\mathrm{L},M},d_{\mathrm{L},M})
\end{pmatrix}
$},
\end{aligned}
\end{equation}
where $\tilde{C}_\ell$ is given in eq.~\eqref{eq:16}. The covariance matrix $\mathcal{C}_\ell$ is the full $(N_{\rm g} + N_{\rm GW}) \times (N_{\rm g} + N_{\rm GW})$ observed angular power spectrum matrix, encompassing galaxy auto-correlation, GW auto-correlation, and their cross-correlation blocks simultaneously. This joint treatment correctly accounts for the correlations among all sub-blocks when computing $\mathcal{C}_\ell^{-1}$, unlike approaches that invert each sub-block independently and sum the resulting Fisher matrices.

The derivatives are evaluated via symmetric finite differences:
\begin{equation}\label{eq:24}
\frac{\partial\mathcal{C}_\ell}{\partial\theta_\alpha} = \frac{\mathcal{C}_\ell(\theta_\alpha+\Delta\theta_\alpha)-\mathcal{C}_\ell(\theta_\alpha-\Delta\theta_\alpha)}{2\Delta\theta_\alpha},
\end{equation}
with a uniform relative step size of $10^{-3}$ for all parameters. For the dark energy EoS parameters $w_0$ and $w_a$, we use absolute step sizes of $\Delta w_0 = \Delta w_a = 10^{-2}$. A small regularization $\epsilon \mathbf{I}$ with $\epsilon = 10^{-10}$ is added to $\mathcal{C}_\ell$ solely during matrix inversion to prevent numerical singularities; this regularization does not enter the finite-difference numerator. We have verified the convergence of all derivatives by varying the step sizes over one order of magnitude.

The full parameter vector in the standard analysis contains 37 parameters: 7 cosmological parameters ($w_0$, $w_a$, $h$, $\Omega_{\rm c}$, $\Omega_{\rm b}$, $n_{\rm s}$, $\ln A_{\rm s}$) together with 15 per-bin galaxy bias scaling factors $\alpha_{\rm g}^{(i)}$ and 15 per-bin GW bias scaling factors $\alpha_{\rm GW}^{(i)}$. When inverting the $37 \times 37$ Fisher matrix, the bias parameters are automatically marginalized over, and the marginalized 1$\sigma$ error on a cosmological parameter $\theta_\alpha$ is $\sigma_{\theta_\alpha} = \sqrt{(F^{-1})_{\alpha\alpha}}$. Note that the matter density $\Omega_{\rm m} = \Omega_{\rm c} + \Omega_{\rm b}$ is not a direct sampling parameter; its uncertainty is derived via error propagation from the covariance matrix, $\sigma^2(\Omega_{\rm m}) = (F^{-1})_{\Omega_{\rm c}\Omega_{\rm c}} + (F^{-1})_{\Omega_{\rm b}\Omega_{\rm b}} + 2(F^{-1})_{\Omega_{\rm c}\Omega_{\rm b}}$.

\section{Modeling the characteristics of the surveys}\label{sec:simulation}

We adopt the CSST photometric survey as the galaxy catalog, chosen for its large sample size (${\sim}1.925\times10^9$ galaxies) and broad redshift coverage ($z \lesssim 4$) \cite{CSST:2025ssq, Gong:2019yxt}. The survey covers approximately 17500~deg$^2$, corresponding to a sky fraction $f_{\rm sky} \approx 0.42$. We model the photometric redshift uncertainty as $\sigma_z = 0.05(1+z)$ and adopt the galaxy number density distribution from ref.~\cite{Gong:2019yxt}. The galaxy sample is divided into 15 redshift bins of equal galaxy counts spanning $z \in [0, 4]$, which also define the luminosity-distance bins for GW sources via the fiducial cosmology.

For the GW event catalog, {we adopt an O1--O3-era BBH population prescription consisting of the Power Law + Peak mass distribution~\cite{Talbot:2018cva}, a Madau-type merger-rate evolution~\cite{Madau:2014bja}, and a reference merger-rate density of $R(z{=}0.2)=23.9~{\rm Gpc^{-3}\,yr^{-1}}$~\cite{KAGRA:2021duu}.}
Parameter estimation uncertainties are computed using a modified version of \texttt{GWFish}\footnote{\url{https://github.com/janosch314/GWFish}} \cite{Dupletsa:2022scg} with the B-DECIGO detector configuration added, supplemented by luminosity-distance errors from peculiar velocities and weak lensing \cite{Song:2022siz}. Events are retained if they satisfy a signal-to-noise ratio (SNR) $\rho > 8$ and $\Delta d_{\mathrm{L}} / d_{\mathrm{L}} < 1$.

\begin{figure*}
    \centering
    \includegraphics[width=0.85\textwidth]{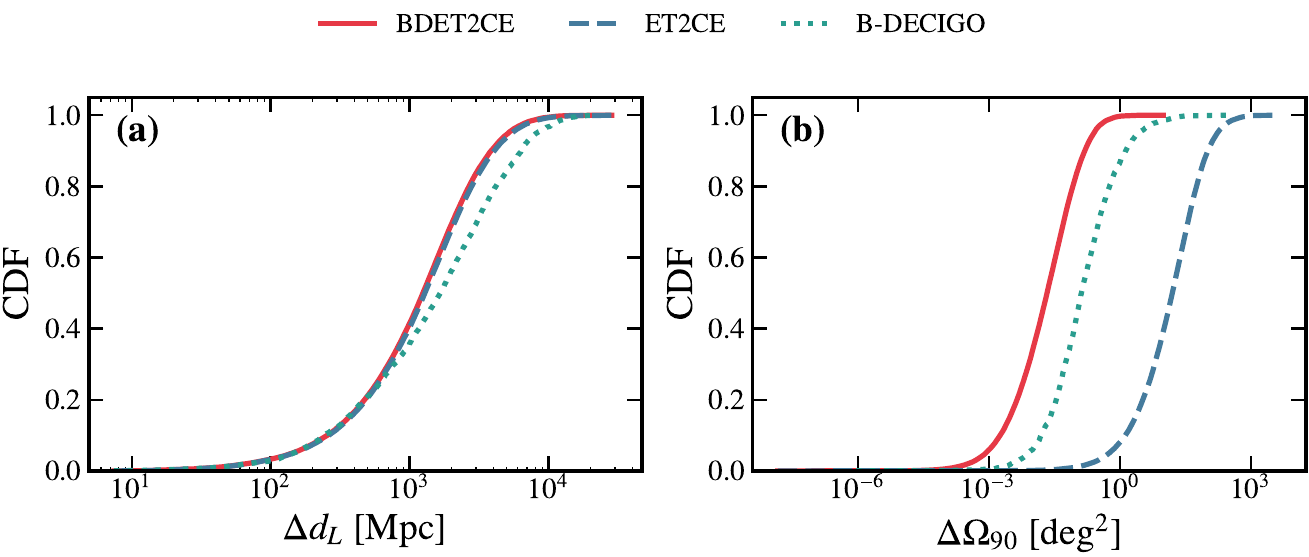}
    \caption{Cumulative distribution functions of the luminosity-distance measurement error $\Delta d_L$ (left) and the 90\% credible-level sky localization area $\Delta\Omega_{90}$ (right) for the three detector configurations. BDET2CE (solid) and ET2CE (dashed) achieve comparable $\Delta d_L$ precision since the ground-based detectors dominate the distance measurement, while B-DECIGO (dotted) yields larger errors due to its lower SNRs. The sky localization shows the opposite hierarchy: BDET2CE achieves median $\Delta\Omega_{90} \approx 0.02$ deg$^2$, roughly three orders of magnitude better than ET2CE ($\sim$16 deg$^2$), demonstrating the transformative advantage of multi-band observation for angular resolution.}
    \label{fig:cdf}
\end{figure*}

We consider three detector configurations to assess the impact of multi-band observations:
\begin{itemize}
    \item \textbf{BDET2CE} (multi-band): B-DECIGO \cite{Kawamura:2020pcg} operating at 0.1--10~Hz combined with ET \cite{Punturo:2010zz} and two CEs \cite{LIGOScientific:2016wof} at 1--$10^4$~Hz. This yields ${\sim}6.46\times10^4$ events per year with a median $\Delta d_{\mathrm{L}}/d_{\mathrm{L}} \approx 9.1\%$ and redshift coverage up to $z \approx 10$.
    \item \textbf{ET2CE} (ground-only): ET + 2CE, providing ${\sim}6.40\times10^4$ events per year with a median $\Delta d_{\mathrm{L}}/d_{\mathrm{L}} \approx 9.4\%$, comparable in event count to BDET2CE but with significantly larger sky localization uncertainties.
    \item \textbf{B-DECIGO} (space-only): B-DECIGO alone, yielding only ${\sim}3.46\times10^3$ events per year with a median $\Delta d_{\mathrm{L}}/d_{\mathrm{L}} \approx 33\%$ and redshift coverage limited to $z \lesssim 3.6$.
\end{itemize}
The key observational properties of the three configurations are summarized in Table~\ref{tab:detectors}. Figure~\ref{fig:cdf} shows the cumulative distribution functions of the luminosity-distance error $\Delta d_L$ and the 90\% credible-level sky localization area $\Delta\Omega_{90}$ for the three configurations. The left panel reveals that BDET2CE and ET2CE achieve comparable $\Delta d_L$ distributions (median $\sim$1250 Mpc), since the luminosity-distance precision is dominated by the ground-based component; B-DECIGO alone yields significantly larger errors (median $\sim$1660 Mpc) due to its lower SNRs. In contrast, the right panel demonstrates the transformative impact of multi-band observation on sky localization: BDET2CE achieves a median $\Delta\Omega_{90} \approx 0.02$ deg$^2$, roughly three orders of magnitude smaller than ET2CE ($\sim$16 deg$^2$), with B-DECIGO at an intermediate level ($\sim$0.13 deg$^2$). This dramatic improvement in sky localization directly translates into a much larger $\ell_{\rm damp}$ for BDET2CE (see Table~\ref{tab:ell}), preserving the angular power spectrum signal to high multipoles and enabling the tighter cosmological and bias constraints presented in sect.~\ref{sec:results}.

\begin{table}
\centering
\caption{Summary of the three GW detector configurations considered in this work. $N_{\rm evt}$ is the number of filtered events per year, $\tilde{\sigma}_{d_{\mathrm{L}}}$ is the median relative luminosity distance error, and $z_{\max}$ is the maximum source redshift.}\label{tab:detectors}
\renewcommand{\arraystretch}{1.3}
\begin{tabular}{lccccc}
\hline\hline
Configuration & Band & $N_{\rm evt}/{\rm yr}$ & $\tilde{\sigma}_{d_{\mathrm{L}}}$ & $z_{\max}$ \\
\hline
BDET2CE & 0.1--$10^4$~Hz & 64632 & 9.1\% & 10.2 \\
ET2CE & 1--$10^4$~Hz & 63997 & 9.4\% & 10.2 \\
B-DECIGO & 0.1--10~Hz & 3460 & 33.2\% & 3.6 \\
\hline\hline
\end{tabular}
\end{table}

\begin{figure*}
    \centering
    \includegraphics[width=0.85\textwidth]{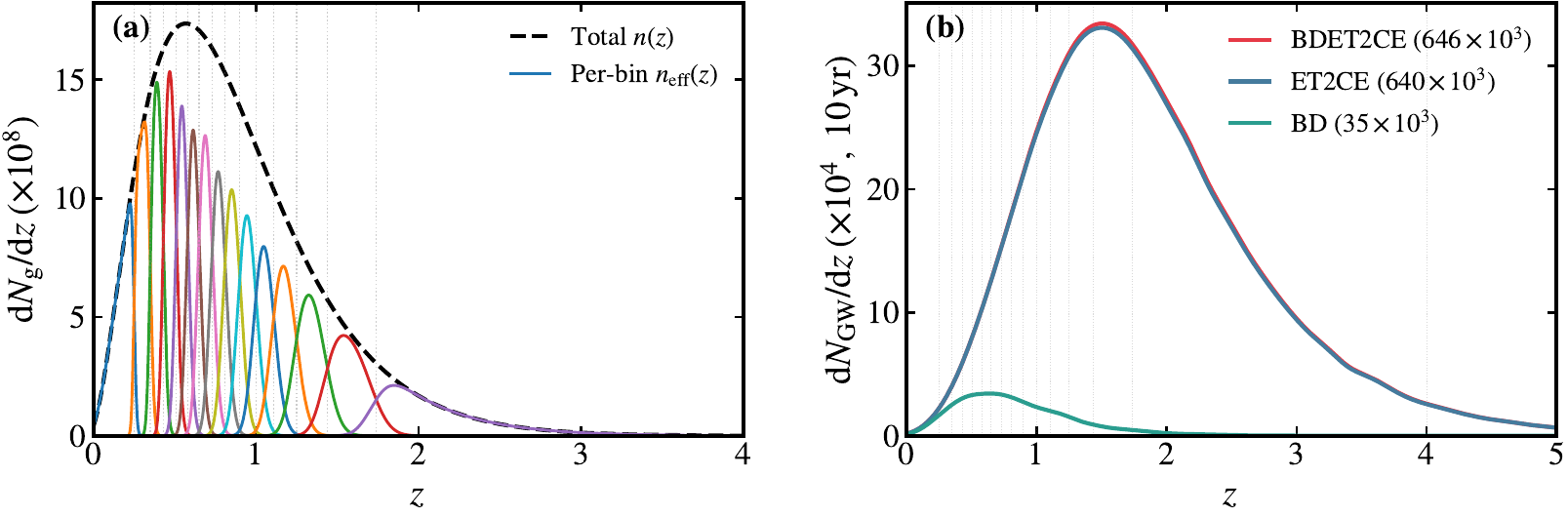}
    \caption{Source redshift and luminosity-distance distributions. \textit{Left:} CSST galaxy number density per unit redshift per steradian (black dashed) and the effective distributions in each of the 15 photometric redshift bins (colored curves), showing the overlap due to photo-$z$ scatter $\sigma_z = 0.05(1+z)$. \textit{Right:} Total GW event rate density as a function of redshift for BDET2CE, ET2CE, and B-DECIGO with $T_{\rm obs} = 10$~yr. BDET2CE and ET2CE have comparable event rates, while B-DECIGO detects far fewer events due to its limited sensitivity to the final coalescence phase of stellar-mass BBH mergers.}
    \label{fig:nz}
\end{figure*}

Figure~\ref{fig:nz} shows the source distributions for the CSST galaxy survey and the three GW detector configurations. The galaxy redshift bins (left panel) are defined by equal number counts, yielding narrower bins at the peak of the $n(z)$ distribution and a broad final bin at $z = 1.74$--$4.0$. The effective per-bin distributions, convolved with the photometric redshift window function, exhibit the characteristic overlap between adjacent bins due to the photo-$z$ scatter $\sigma_z = 0.05(1+z)$. The GW per-bin distributions are similarly convolved with the luminosity-distance measurement uncertainty through the window function in eq.~\eqref{eq:9}, producing analogous inter-bin overlap in $d_L$ space. The right panel shows the total GW event rate distributions for the three detector configurations with 10 years of observations. BDET2CE and ET2CE yield comparable event rates ($\sim 6.5 \times 10^4$ yr$^{-1}$) since both include the ground-based ET + 2CE network, which dominates the detection count; B-DECIGO alone detects far fewer events ($\sim 3.5 \times 10^3$ yr$^{-1}$) because its deci-hertz sensitivity band captures only the early inspiral phase of stellar-mass BBH mergers, where the accumulated SNR is relatively low.

\section{Results and discussions}\label{sec:results}

We present the Fisher forecast results for the angular power spectrum cross-correlation between the CSST photometric galaxy survey and GW events from the three detector configurations described in sect.~\ref{sec:simulation}. The fiducial cosmological parameters follow the Planck 2018 results: $h = 0.6766$, $\Omega_{\rm c} = 0.26069$, $\Omega_{\rm b} = 0.04897$, $n_{\rm s} = 0.9665$, $A_{\rm s} = 2.1\times10^{-9}$, $w_0 = -1$, $w_a = 0$. In the standard analysis, we adopt $\ln(10^{10}A_{\rm s})$ for numerical stability and employ per-bin bias scaling factors as described in sect.~\ref{ssec:multi-tracer}.

\subsection{Cosmological parameter constraints}\label{ssec:cosmological}

Table~\ref{tab:cosmo} summarizes the cosmological parameter constraints from the joint galaxy$\times$GW analysis in both the $\Lambda$CDM and $w_0w_a$CDM frameworks, alongside the galaxy-only baseline obtained from the $15\times15$ galaxy auto-correlation block alone.

In the $\Lambda$CDM model, the cross-correlation yields dramatic improvements on the Hubble constant. With 10 years of observations, BDET2CE achieves $\sigma(h)/h = 0.35\%$, ET2CE achieves $0.55\%$, and B-DECIGO achieves $2.45\%$, all substantially tighter than the galaxy-only baseline of $3.87\%$. The multi-band advantage is pronounced: BDET2CE improves $\sigma(h)/h$ by 37\% over the ground-only ET2CE and by 86\% over B-DECIGO alone. This hierarchy reflects the interplay between event rate and sky localization: BDET2CE and ET2CE share comparable event catalogs, but the multi-band sky localization preserves the cross-correlation signal to higher multipoles; B-DECIGO has excellent localization but is limited by its much smaller event rate. Even with only 1 year of observations, BDET2CE already achieves $\sigma(h)/h = 0.70\%$, while B-DECIGO requires the full 10 years to reach $2.45\%$. The constraint on $\Omega_{\rm m}$ also improves, though more modestly.

Extending to the $w_0w_a$CDM framework, the additional degeneracies between $h$ and the dark energy EoS parameters weaken all constraints (see Table~\ref{tab:cosmo} for the full results). The GW cross-correlation still improves $\sigma(h)/h$ substantially (from $6.22\%$ to $3.82\%$ for BDET2CE at 10~yr, a $\sim$39\% gain), but the improvement on the dark energy EoS is more moderate ($\sim$14\% for $w_0$ and $\sim$15\% for $w_a$). This contrast reveals a key physical insight: the cross-correlation is most powerful for constraining the distance scale $h$, which directly enters the distance-redshift mapping, while its leverage on the dark energy EoS is limited by parameter degeneracies. The detector hierarchy follows the same pattern as in $\Lambda$CDM, but the multi-band advantage becomes more moderate ($\sim$4\% improvement over ET2CE, compared with 37\% in $\Lambda$CDM), because the dark energy constraints are driven primarily by low multipoles where both configurations retain adequate signal.

\begin{figure*}
    \centering
    \includegraphics[width=0.95\textwidth]{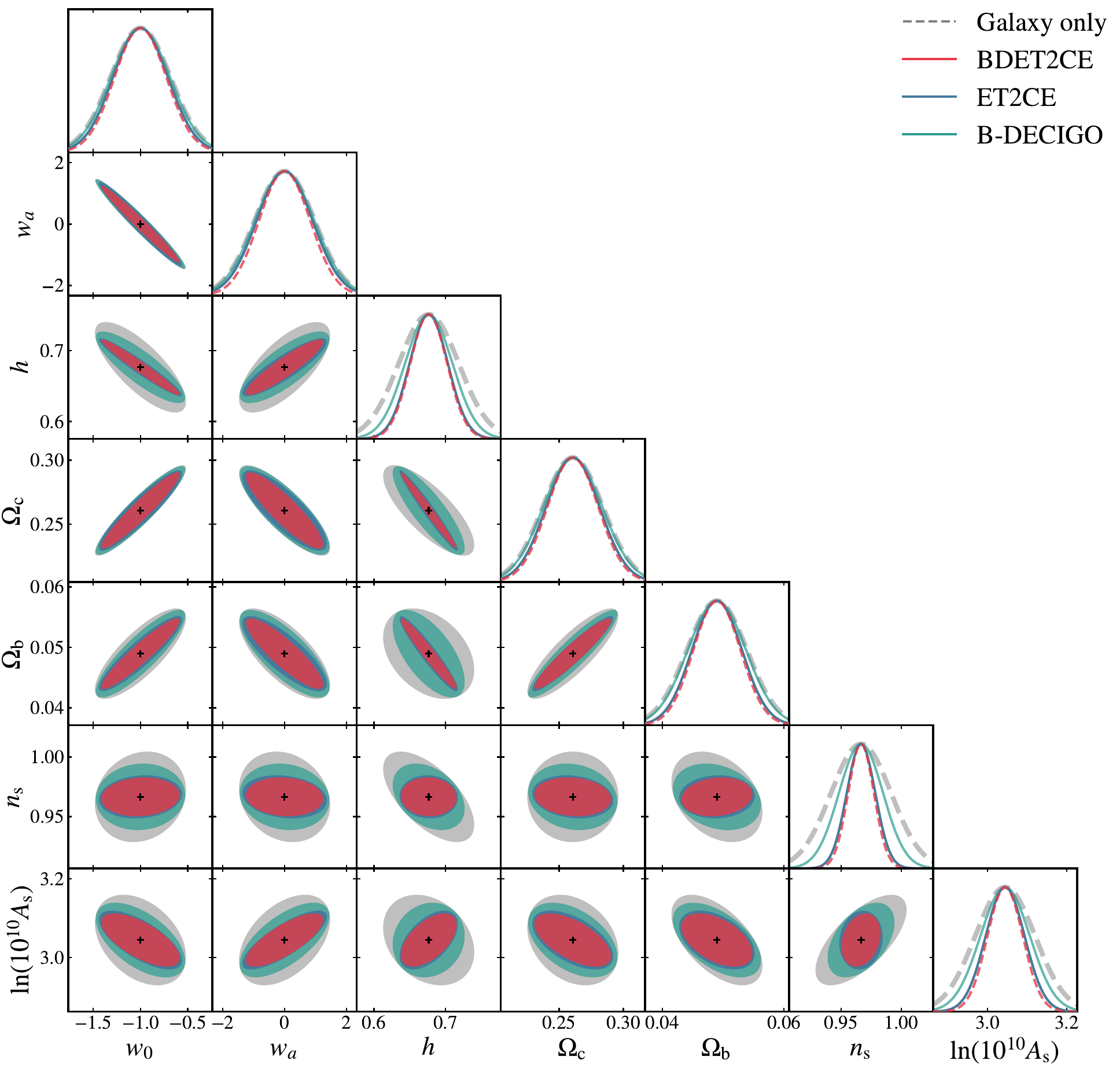}
    \caption{Fisher forecast $1\sigma$ contours for the seven cosmological parameters in the $w_0w_a$CDM model with $T_{\rm obs} = 10$~yr. Gray dashed ellipses show the galaxy-only baseline (15$\times$15 auto-correlation block); colored ellipses show the joint analysis with GW cross-correlation for BDET2CE (red), ET2CE (blue), and B-DECIGO (green). The diagonal panels display the marginalized one-dimensional distributions.}
    \label{fig:corner}
\end{figure*}

Figure~\ref{fig:corner} displays the two-dimensional marginalized Fisher contours for all seven cosmological parameters, comparing the galaxy-only case (gray dashed) with the three joint analyses. The most significant improvement appears in $h$, where the contours shrink dramatically, consistent with the direct sensitivity of the cross-correlation to the distance scale. The parameters $n_{\rm s}$ and $\ln(10^{10}A_{\rm s})$ also show visible improvement, because the cross-correlation introduces an independent tracer combination that helps break the degeneracy between the power spectrum amplitude and the clustering biases present in the galaxy auto-correlation alone. Across all parameters, the detector hierarchy BDET2CE $>$ ET2CE $>$ B-DECIGO is maintained.

\begin{figure*}
    \centering
    \includegraphics[width=0.85\textwidth]{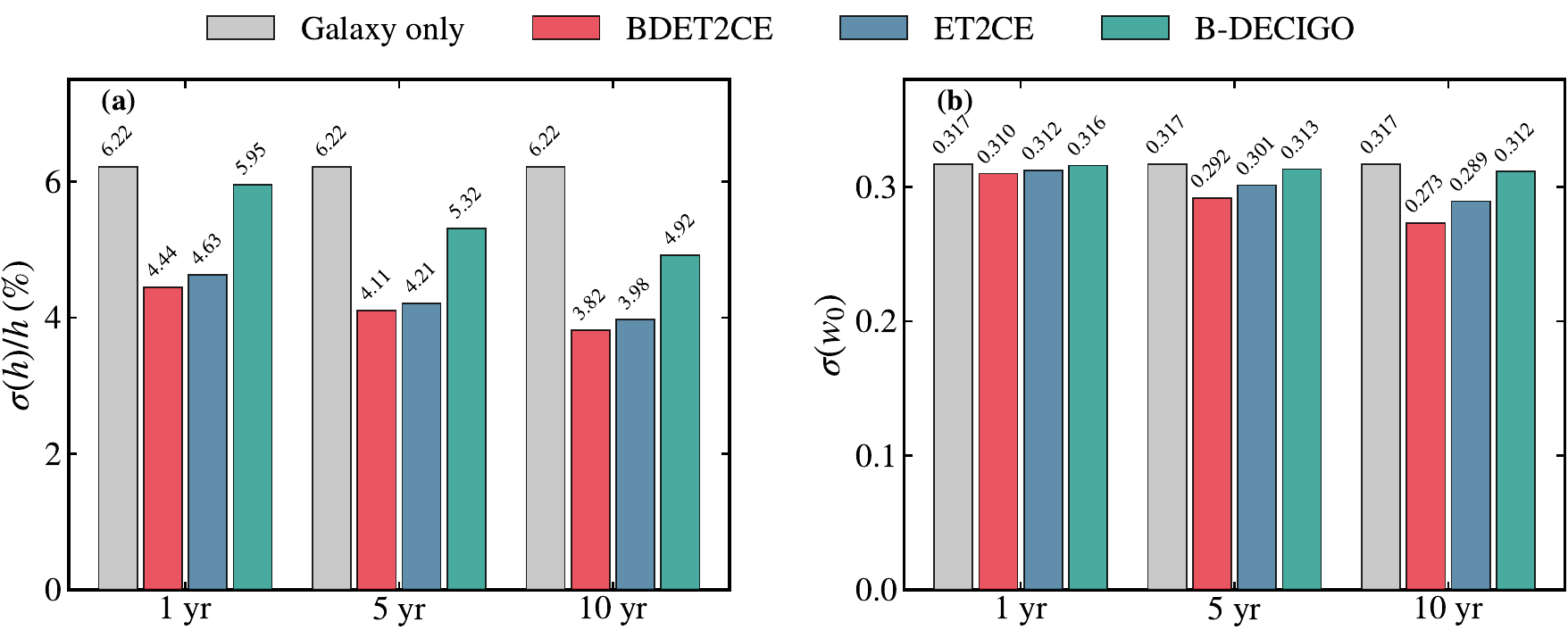}
    \caption{Marginalized $1\sigma$ constraints on $h$ (left) and $w_0$ (right) in the $w_0w_a$CDM model as a function of GW observation duration for three detector configurations. The gray bars show the galaxy-only baseline, which is independent of $T_{\rm obs}$. The joint constraints tighten monotonically with increasing GW observation time, with BDET2CE achieving the largest improvement.}
    \label{fig:detcomp}
\end{figure*}

The observation time dependence in the $w_0w_a$CDM framework further reveals the origin of the improvement. Figure~\ref{fig:detcomp} compares the $1\sigma$ constraints on $h$ and $w_0$ across different detector configurations and GW observation durations. The galaxy-only baseline is independent of $T_{\rm obs}$ (since it depends only on the CSST survey), while the joint constraints tighten monotonically with increasing GW observation duration. For BDET2CE, $\sigma(h)/h$ improves from 4.44\% (1 yr) to 3.82\% (10 yr), a 14\% gain driven by the accumulation of GW events and the resulting increase in cross-correlation SNR.

\red{\begin{table*}
\centering
\caption{Marginalized $1\sigma$ constraints on cosmological parameters. The upper block shows $\Lambda$CDM results (5 cosmological + 15 galaxy bias + 15 GW bias = 35 free parameters for joint; 5 cosmological + 15 galaxy bias = 20 for galaxy only). The lower block shows $w_0w_a$CDM results (7 + 15 + 15 = 37 for joint; 7 + 15 = 22 for galaxy only). The ``Galaxy only'' column uses the $15\times 15$ galaxy auto-correlation block alone. The ``Joint'' columns use the full $30\times 30$ covariance matrix including galaxy auto, GW auto, and cross-correlations.}\label{tab:cosmo}
\renewcommand{\arraystretch}{1.4}
\begin{tabular*}{\textwidth}{@{\extracolsep{\fill}}lccccccccccc}
\hline\hline
 & Galaxy & \multicolumn{3}{c}{BDET2CE (Joint)} & \multicolumn{3}{c}{ET2CE (Joint)} & \multicolumn{3}{c}{B-DECIGO (Joint)} \\
\cmidrule(lr){3-5} \cmidrule(lr){6-8} \cmidrule(lr){9-11}
Parameter & only & 1 yr & 5 yr & 10 yr & 1 yr & 5 yr & 10 yr & 1 yr & 5 yr & 10 yr \\
\hline
\multicolumn{11}{c}{$\Lambda$CDM} \\
\hline
$\sigma(h)/h$ & 3.87\% & 0.70\% & 0.41\% & 0.35\% & 1.49\% & 0.74\% & 0.55\% & 3.61\% & 2.93\% & 2.45\% \\
$\sigma(\Omega_{\rm m})/\Omega_{\rm m}$ & 1.83\% & 1.62\% & 1.60\% & 1.57\% & 1.66\% & 1.64\% & 1.63\% & 1.81\% & 1.75\% & 1.72\% \\
\hline
\multicolumn{11}{c}{$w_0w_a$CDM} \\
\hline
$\sigma(w_0)$ & 0.317 & 0.310 & 0.292 & 0.273 & 0.312 & 0.301 & 0.290 & 0.316 & 0.313 & 0.312 \\
$\sigma(w_a)$ & 0.970 & 0.945 & 0.885 & 0.825 & 0.958 & 0.932 & 0.902 & 0.968 & 0.962 & 0.957 \\
$\sigma(h)/h$ & 6.22\% & 4.44\% & 4.11\% & 3.82\% & 4.63\% & 4.21\% & 3.98\% & 5.95\% & 5.32\% & 4.92\% \\
$\sigma(\Omega_{\rm m})/\Omega_{\rm m}$ & 8.91\% & 8.73\% & 8.12\% & 7.55\% & 8.80\% & 8.40\% & 7.96\% & 8.90\% & 8.85\% & 8.81\% \\
\hline\hline
\end{tabular*}
\end{table*}}

Several studies have forecast CSST cosmological constraints including cosmic shear in $3\times 2$pt analyses \cite{Gong:2019yxt, Xiong:2024dtx, Lin:2022aro}, and ref.~\cite{Su:2025zuc} further combined CSST cosmic shear with GW clustering, achieving $\sigma(h)/h \approx 2.2\%$. Our analysis focuses on the clustering-only channel without cosmic shear or external priors, isolating the contribution of the GW$\times$galaxy cross-correlation alone. Ref.~\cite{Pedrotti:2025tfg} forecast the cross-correlation of GW events from 2 ET + 2 CE with the Euclid photometric survey in flat $\Lambda$CDM, reporting a joint constraint of $\sigma(H_0)/H_0 \sim 1\%$ with a galaxy-only baseline of $\sim$15\%. The larger improvement factor ($\sim$15 vs.\ our $\sim$11 for BDET2CE) reflects their weaker galaxy-only baseline due to a more conservative $\ell_{\max}$ cut. Importantly, even with the strong CSST galaxy-only baseline ($3.87\%$ in $\Lambda$CDM), the GW cross-correlation provides additional independent geometric information through the $d_{\mathrm{L}}(z)$ mapping, pushing the constraint to $0.35\%$ for BDET2CE. This additional information is qualitatively distinct from what galaxy clustering alone can provide, as it probes the absolute distance scale rather than the shape of the power spectrum. The tighter absolute constraint compared to ref.~\cite{Pedrotti:2025tfg} is primarily driven by the multi-band sky localization, which preserves the cross-correlation signal to higher multipoles.

\subsection{GW clustering bias constraints}\label{ssec:gwbias}

\begin{figure}
    \centering
    \includegraphics[width=\columnwidth]{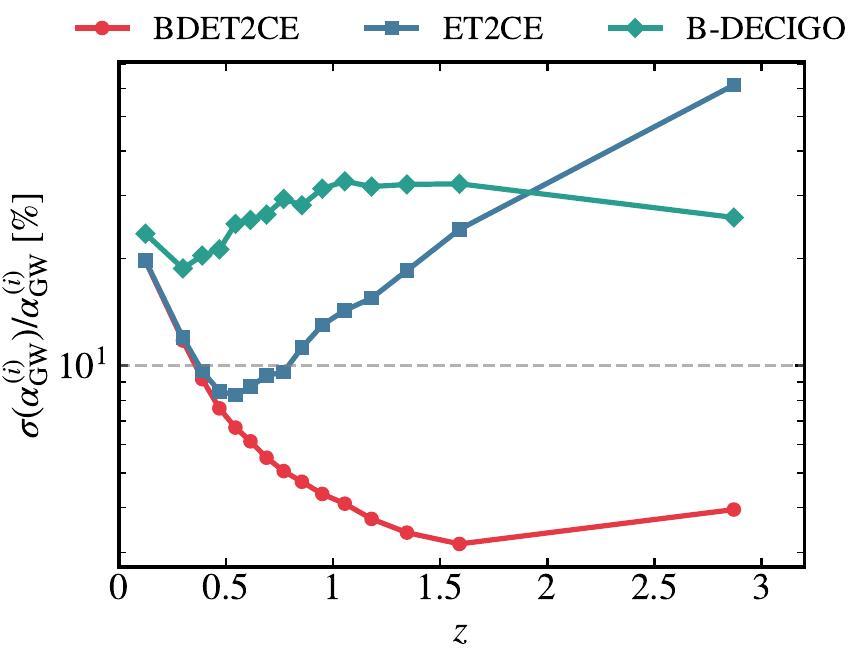}
    \caption{Per-bin GW clustering bias fractional uncertainties $\sigma(\alpha_{\rm GW}^{(i)})/\alpha_{\rm GW}^{(i)}$ for 10 years of observation with three detector configurations. BDET2CE (red) achieves the tightest constraints ($\sim$3\% at $z \sim 1.3$--$1.7$), while ET2CE (blue) degrades sharply at $z > 0.5$ due to its poor sky localization ($\ell_{\rm damp} \sim 40$--$900$). B-DECIGO (green) yields constraints of $\sim$20--33\% limited by its smaller event catalog.}
    \label{fig:gwbias}
\end{figure}

A distinctive capability of the cross-correlation is the per-bin measurement of the GW clustering bias $b_{\rm GW}(z)$, which carries astrophysical information about compact binary formation channels and host environments. Figure~\ref{fig:gwbias} shows the fractional uncertainties $\sigma(\alpha_{\rm GW}^{(i)})/\alpha_{\rm GW}^{(i)}$ for each luminosity-distance bin with $T_{\rm obs} = 10$~yr.

For BDET2CE, the bias is constrained to $\sim$3\% precision at $z \sim 1.3$--$1.7$ where the GW event density peaks, degrading to $\sim$20\% at the lowest redshifts ($z < 0.25$) where the comoving volume is small, and to $\sim$4\% at $z \sim 3$ for the widest high-redshift bin. The three detector configurations show strikingly different behavior. BDET2CE achieves the tightest constraints across all redshifts, benefiting from both a high event rate and excellent sky localization that preserves the angular power spectrum signal to high multipoles. ET2CE, despite having a comparable event rate ($\sim$64000 yr$^{-1}$), yields significantly weaker bias constraints at $z > 0.5$, reaching $\sim$60\% at $z \sim 3$. This degradation is a direct consequence of the poorer sky localization of the ground-only network ($\ell_{\rm damp} \sim 40$--$900$ for ET2CE vs.\ $\sim$1600--31000 for BDET2CE), which strongly suppresses the GW angular power spectrum at high multipoles and limits the information available for bias measurement. B-DECIGO, with its superior localization but limited event rate ($\sim$3500 yr$^{-1}$), yields constraints of $\sim$20--33\%.

This result highlights a qualitative difference between the cosmological and astrophysical returns of multi-band observation. For cosmological constraints, the improvement from BDET2CE over ET2CE is modest ($\sim$4\% in $\sigma(h)/h$), because the distance-redshift mapping is probed primarily at low multipoles where both configurations retain adequate signal. For GW bias measurements, in contrast, the multi-band localization advantage is decisive: it preserves the angular power spectrum signal to high $\ell$, providing the additional modes needed for precise per-bin bias recovery.

Refs.~\cite{Raccanelli:2016cud, Scelfo:2018sny, Scelfo:2020jyw} studied the GW$\times$galaxy cross-correlation as a probe of compact binary formation channels and predicted a cross-correlation SNR of order a few for ET. Ref.~\cite{Libanore:2020fim} demonstrated that the GW clustering bias can be detected at high statistical significance with next-generation detectors. Ref.~\cite{Su:2025zuc} reported GW bias constraints of $\sim$4--5\% per bin using four broader bins and CSST cosmic shear; our finer 15-bin scheme resolves the bias evolution at higher redshift resolution. The redshift profile of $\sigma(\alpha_{\rm GW}^{(i)})/\alpha_{\rm GW}^{(i)}$ traces the convolution of the GW event rate, sky localization quality, and survey geometry, and future measurements could distinguish among compact binary formation channels that predict distinct bias evolutions \cite{Scelfo:2020jyw, Chakravarti:2026bcv, Peron:2023zae}. For example, stellar-origin BBH mergers trace their host galaxies and are expected to exhibit a clustering bias that increases with redshift, whereas a primordial black hole population would track the dark matter distribution with $b_{\rm GW} \sim 1$ roughly independent of redshift \cite{Raccanelli:2016cud, Libanore:2023ovr}; the percent-level, redshift-resolved bias measurements achievable with BDET2CE could probe such differences.

\subsection{Limitations and outlook}\label{ssec:outlook}

Several limitations of the present analysis should be noted. {First, our fiducial GW catalog adopts an O1--O3-era BBH population prescription. We estimate the impact of this assumption by reweighting the catalog to the GWTC-5.0 population distribution and rescaling the event number to $R(z=0.2)=30.4~{\rm Gpc^{-3}\,yr^{-1}}$~\cite{LIGOScientific:2026ctl}, while retaining the fiducial single-event distance and localization errors. For the 10-year BDET2CE configuration, this update tightens the constraints on $h$, $w_0$, and $w_a$ by approximately $3\%$. The relative changes in the GW-bias uncertainties range from $3.6\%$ to $16\%$, while the best-constrained bin remains at the $3\%$ level, with an uncertainty of $3.28\%$ compared with $3.16\%$ in the baseline. The tested population update therefore does not alter our main conclusions.} Second, we do not include photometric redshift bias parameters $\Delta z^{(i)}$, which represent systematic shifts in the photometric redshift calibration; such biases would degrade the bin-overlap sensitivity that drives the cosmological constraints, and our omission means the forecasted errors may be somewhat optimistic. In addition, the Fisher matrix formalism assumes Gaussian posteriors and cannot capture non-Gaussian degeneracies that may arise in the full parameter space.

Several directions could substantially improve the constraints in future work. Incorporating external priors from Planck CMB observations would tighten the constraints on $h$, $\Omega_{\rm b}$, and $n_{\rm s}$, breaking degeneracies that currently limit the dark energy EoS measurement \cite{Guo:2018gyo,Li:2024qso, Li:2025owk, Li:2025dwz, Li:2025eqh, Du:2025iow, Du:2025xes, Feng:2025mlo, Wu:2025vfs, Li:2025htp, Ling:2025lmw, Li:2025muv, Zhang:2025dwu}. Extending the analysis to a $3\times 2$pt framework by including cosmic shear \cite{Xiong:2024dtx, Su:2025zuc} would further sharpen the constraints. On the GW side, independent calibration of the clustering bias from numerical simulations \cite{Peron:2023zae} or multi-messenger observations would convert $b_{\rm GW}(z)$ from a nuisance parameter into a physically informative quantity for probing compact binary formation channels. The cross-correlation method could also be combined with spectral sirens \cite{Cheng:2026atn}, which encode complementary information from the GW mass distribution and merger rate evolution rather than from the spatial clustering exploited here, enabling a joint inference that leverages the full dark siren information content. Finally, spectroscopic galaxy surveys with precise redshift measurements would reduce the photo-$z$ scatter that smears the bin overlap, and extending the analysis to other GW source types such as binary neutron star and neutron star-black hole mergers would provide additional tracers with distinct bias evolution and redshift coverage.

\section{Conclusions}\label{sec:conclusions}

In this work, we have performed a Fisher forecast for the angular power spectrum cross-correlation between multi-band GW observations and the CSST photometric galaxy survey in both the $\Lambda$CDM and $w_0w_a$CDM frameworks. By binning GW events in luminosity-distance space and galaxy sources in photometric redshift space, the cross-correlation directly probes the distance-redshift relation $d_{\mathrm{L}}(z)$, providing a geometric lever arm for constraining cosmological parameters. We compared three detector configurations: the multi-band BDET2CE (B-DECIGO + ET + 2CE), the ground-only ET2CE, and the space-only B-DECIGO.

Our main findings are as follows.

(i) The GW$\times$galaxy cross-correlation provides substantial improvements on the Hubble constant. In the $\Lambda$CDM model, the multi-band BDET2CE achieves $\sigma(h)/h = 0.35\%$ with 10 years of observations, tightening the constraint by 37\% relative to the ground-only ET2CE ($0.55\%$) and by 86\% relative to B-DECIGO alone ($2.45\%$). In the $w_0w_a$CDM framework, the additional degeneracies with the dark energy EoS parameters weaken all constraints: BDET2CE achieves $\sigma(h)/h = 3.82\%$, with more moderate multi-band improvements of $\sim$4\% over ET2CE and $\sim$22\% over B-DECIGO.

(ii) The most striking multi-band advantage lies in the per-bin measurement of the GW clustering bias. At $z \sim 1$--$2$, BDET2CE constrains the bias to $\sim$3\% precision, compared to $\sim$8--60\% for ET2CE and $\sim$20--33\% for B-DECIGO. This contrast arises because bias recovery requires high-$\ell$ modes that are preserved by the multi-band sky localization but severely damped in the ground-only configuration. These precise, redshift-resolved measurements open a new avenue for probing the astrophysics of compact binary mergers.

These results demonstrate the complementary roles of galaxy surveys and GW observations in cross-correlation cosmology. The CSST galaxy auto-correlation provides the dominant constraining power on cosmological parameters, while the GW cross-correlation contributes unique geometric information through the $d_{\mathrm{L}}(z)$ relation and enables per-bin measurements of the GW clustering bias. Multi-band observation is essential for realizing the full potential of this synergy: the combination of high event rates from ground-based detectors with the superior sky localization of space-borne instruments yields improvements that neither component can achieve independently.

\begin{acknowledgments}
    
This work was supported by the National Natural Science Foundation of China 
(Grants Nos. 12473001, 12575049, and 12533001), the National SKA Program of China 
(Grants Nos. 2022SKA0110200 and 2022SKA0110203), the China Manned Space Program 
(Grant No. CMS-CSST-2025-A02), and the 111 Project (Grant No. B16009).

\end{acknowledgments}


\bibliography{ref}

\end{document}